\documentclass[pdftex, twocolappendix, numberedappendix, iop, apj]{emulateapj}
\usepackage{graphicx}
\usepackage{dcolumn}
\usepackage{bm}
\usepackage{amssymb}
\usepackage{latexsym}
\usepackage{booktabs}
\usepackage{amsmath}
\usepackage{multirow}
\usepackage[colorlinks=true, linkcolor=red, citecolor=blue]{hyperref}
\newcommand{\be}{\begin{equation}}
\newcommand{\ee}{\end{equation}}
\newcommand{\bq}{\begin{eqnarray}}
\newcommand{\eq}{\end{eqnarray}}
\newcommand{\lcdm}{\ensuremath{\Lambda\mathrm{CDM}}}
\newcommand{\planck}{\ensuremath{\it{Planck}}}
\newcommand{\WMAP}{\ensuremath{\it{WMAP}}}
\newcommand{\wmap}{\ensuremath{\it{WMAP}}}
\begin{document}

\title{Assessing Consistency Between \WMAP{} 9-year and \planck{} 2015 Temperature Power Spectra}

\author{Y.~Huang\altaffilmark{1}, G.~E.~Addison\altaffilmark{1}, J.~L.~Weiland\altaffilmark{1},  C.~L.~Bennett\altaffilmark{1}}

\email{yhuang98@jhu.edu}

\altaffiltext{1}{
Dept. of Physics \& Astronomy, The Johns Hopkins University, 3400 N. Charles St., Baltimore, MD 21218-2686
}
\begin{abstract}
We perform a comparison of \WMAP{} 9-year (\WMAP{}9) and \planck{} 2015 cosmic microwave background (CMB) temperature power spectra across multipoles $30\leq\ell\leq1200$. We generate simulations to estimate the correlation between the two datasets due to cosmic variance from observing the same sky. We find that their spectra are consistent within $1\sigma$. While we do not implement the optimal ``$C^{-1}$" estimator on \WMAP{} maps as in the \WMAP{}9 analysis, we demonstrate that the change of pixel weighting only shifts our results at most at the $0.66\sigma$ level. We also show that changing the fiducial power spectrum for simulations only impacts the comparison at around $0.1\sigma$ level. We exclude $\ell<30$ both because \wmap9 data were included in the \planck\ 2015 $\ell<30$ analysis, and because the cosmic variance uncertainty on these scales is large enough that any remaining systematic difference between the experiments is extremely unlikely to affect cosmological constraints. The consistency shown in our analysis provides high confidence in both the \WMAP{}9 temperature power spectrum and the overlapping multipole region of \planck{} 2015's, virtually independent of any assumed cosmological model. Our results indicate that cosmological model differences between \planck{} and \WMAP{} do not arise from measurement differences, but from the high multipoles not measured by \WMAP{}.

\end{abstract}

 \maketitle

\section{Introduction}
Observations of the cosmic microwave background (CMB) temperature anisotropy and 
its power spectrum (hereafter TT spectrum) have provided great insight into early universe physics and enabled precise constraints on cosmological parameters, within the context of the $\Lambda$ cold dark matter (\lcdm{}) model \citep[e.g.,][]{Bennett2013,planck/13:2015,sievers/etal:2013,story/etal:2013}. The importance of the determination of cosmological parameters goes beyond studies involving the CMB. The choice of which CMB data set to use can meaningfully impact results from dark matter or hydrodynamic simulations, with implications for constraints on neutrino mass and alternative gravity models \citep[e.g.,][]{Hojiati2015,McCarthy2018,planckXIV}.

 

In recent years, tensions have been shown to exist between CMB and several low-redshift, late time observations as well as within CMB measurements. For example, the most recent constraint on the Hubble constant, $H_0$, from \cite{riess2018} yielded a value $3.7\sigma$ higher than that from \planck{} 2015 \citep{planck/intermediate/46:2016}. Moreover, a $2-3\sigma$ difference was shown between \planck{} 2015 and measurements of weak gravitational lensing \citep[e.g.,][]{Joudaki2017,Kohlinger2017}, concerning the parameter combination $S_8$\footnote{$\bm{S_8}\equiv \sigma_8(\Omega_m/0.3)^{
0.5}$, where $\sigma_8$ is the present day amplitude of the matter fluctuation spectrum and $\Omega_m$ is the present day matter density in units of the critical density.}, which describes the growth of cosmic structure. In addition, a $2.5\sigma$ discordance has been reported between the \planck{} 2015 $\ell < 1000$ and $1000\leq\ell\leq 2508$ data in $\Omega_c h^2$, the cold dark matter density, which is a parameter highly correlated with those constrained by low-redshift measurement \citep{addison2016}. However, using simulated \planck-like TT power spectra, \cite{planck/intermediate/08:2016} argued that these $\ell$-range related shifts in parameters were not statistically significant across the full \lcdm{} model space.

Given the importance of CMB constraints for current and future cosmology, and the existing tensions, it is crucial that CMB measurements are scrutinized.
The two latest full-sky CMB surveys, \WMAP{} \citep{Bennett2013} and \planck{} \citep{planck/01:2015}, provide a valuable opportunity for consistency checks. 

The difference between the \planck{} 2015 and \WMAP{}9 spectra is within the \WMAP{}9 uncertainties  \citep{planck/11:2015}, and the value of each of the \lcdm{} parameters is consistent within 1.5 times the \WMAP{}9 uncertainty \citep{planck/01:2015}. However, we should note that 
the correlation between the two experiments is not negligible as they measure the same sky. Between the \WMAP{}9 and the \planck{} 2013 release \citep{planck/16:2013}, \cite{larson2015} found a $\sim 6\sigma$ parameter difference, with a minimal \lcdm{} model assumed and the correlation between the two experiments accounted for. The \planck{} calibration was significantly revised in the 2015 release \citep{planck/8:2016}, and this motivates revisiting the comparison with \wmap.



This paper therefore investigates consistency between \wmap9 and \planck\ 2015 TT spectra and estimates their correlation using simulations. We examine the multipole range common to both experiments where power-spectrum based likelihoods are employed ($30\leq\ell\leq1200$).
Only when their correlation is quantified, can we quantify their agreement/disagreement in a meaningful sense. Unlike comparisons of parameters, comparisons of power spectra are minimally dependent on the assumed model. We will show in Section 4 that the choice of fiducial model used to generate simulated spectra and estimate covariance between \WMAP{}9 and \planck{} 2015 has a negligible effect on our results. Thus discrepancies that appear between the power spectra would be an indication of experimental systematic errors, instead of evidence for physics beyond the standard model of cosmology.  




We exclude the $\ell<30$ region of the TT spectra from our analysis. Comparison of $\ell<30$ spectra is complicated by the fact that the \planck\ 2015 results included \wmap9 data in a multifrequency fit. A different (pixel-based) likelihood and treatment of foregrounds is also required for these scales. The \planck\ 2015 and \wmap9 results for $\ell<30$ are shown in Figure~2 of \cite{planck/11:2015} and agree within a small fraction of the uncertainty for most of the multipoles. Differences due to imperfect noise or foreground modeling, or some other systematic error, could still exist. Given the size of the cosmic variance uncertainty at $\ell<30$, however, it seems highly unlikely that they could meaningfully impact cosmological results.

Following the completion of this work, the \planck{} team has released their latest results \citep{planck2018I}. The 2015 and 2018 \planck{} TT spectra are in good agreement, as described in Section~3.6 of \cite{planck2018VI}.  We therefore expect the level of consistency between \wmap{}9 and \planck{} 2018 to remain the same.

The outline of this paper is as follows. We describe our simulation procedures in Section 2 and test simulation fidelity in Section 3. We present results in Section 4, followed by conclusions in Section 5.

\section{Simulating TT spectra and Covariance}


Our goal is to quantify the consistency between the TT power spectra observed by \WMAP{}9 and \planck{} 2015. Both teams provide estimates of their experiment's power spectrum $C_{\ell}-C_{\ell'}$ covariance matrices, however we also need the  \WMAP{}9 $\times$ \planck{} 2015 cross-covariance due to common cosmic variance. To estimate this we generated 4000 full-sky simulations of CMB temperature fluctuations. The outline of the simulation procedure is as follows.

\begin{table}\small
\setlength\tabcolsep{6pt}

\begin{center}
\renewcommand{\arraystretch}{1.5}
\begin{tabular}{cccccccccc}
\\
\hline\hline
Parameter  & \WMAP{}9 TT & \planck{} 2015 TT \\ \hline
\multicolumn{3}{c}{\textbf{Fit \lcdm{} parameters}} \\[0.2mm]
$\Omega_bh^2$      & 0.02230
                   & 0.02215

                   \\

$\Omega_ch^2$      &0.1158
                   & 0.1215

                   \\

$100\theta_{MC}$                & 1.0393 & 1.0405

                   \\
$\tau$                & 0.07 & 0.07

                   \\

$\log A_s$           & 3.059
                   & 3.078

                   \\

$n_s$       & 0.9615
                   & 0.9595
                   \\  \hline
 \multicolumn{3}{c}{\textbf{Derived parameters}} \\[0.2mm]
$H_0$[km s$\bm{^{-1}}$Mpc$\bm{^{-1}]}$       & 68.24
                   & 66.52
                   \\
$\Omega_m$       & 0.2981
                   & 0.3261
                   \\
$\sigma_8$       & 0.8007
                   & 0.8288
                   \\
\hline

\end{tabular}
\end{center}
\caption{The cosmological parameters describing the fiducial models used in our simulations. The second column shows the best-fit model from the $30 \leq \ell\leq 1200$ \WMAP{}9 TT spectrum and the third that of the $30 \leq \ell\leq 2508$ \planck{} 2015 spectrum. They are both results of running the best-fit finding algorithm in CosmoMC \citep{cosmomc} with $\tau$ fixed to be 0.07, together with the PICO code \citep{pico}, which computes CMB power spectra given the model.\label{param}}
\end{table}

\begin{enumerate}
\item Generate a set of spherical harmonic coefficients $a_{\ell m}$ using the \texttt{sphtfunc.synalm} routine in \texttt{Healpy}\footnote{A Python implementation of \texttt{Healpix} \citep{Healpix}, see \url{https://healpy.readthedocs.io/en/latest/}.} from a fiducial TT power spectrum chosen to be the best-fit model from the $\ell \geq 30$ \WMAP{}9 TT spectrum. See Table \ref{param} for the cosmological parameters in this model. Unless otherwise noted, results shown come from this fiducial model. Table \ref{param} also includes an alternative model, which is one from the $\ell \geq 30$ \planck{} 2015 TT spectrum. We will use the alternative model to test the stability of our results against different input, see Section 2.3 and Section 4. We also note that this is the only place where an assumed cosmological model comes in.
\item Multiply the $a_{\ell m}$ coefficients with the appropriate beam and pixel functions, then convert them into a CMB map using the \texttt{sphtfunc.alm2map} routine in \texttt{Healpy}.
 \item  Add to the map white noise with variance given by the experiments.
 \item  Apply sky masks and compute the TT spectrum 
 from the masked maps using \texttt{PolSpice}\footnote{\url{http://www2.iap.fr/users/hivon/software/PolSpice/}} \citep{PolSpice1,PolSpice2}.
 \item Compute analytically the power spectrum covariance (hereafter referred to as the analytic covariance), following the prescriptions given by Appendix C1.1.1 of \cite{planck/11:2015}. The analytic calculations take into account the effects of masking, beam and pixel window functions, and the instrumental noise \citep{myth}. We will refer to this approach as ``MASTER" \citep{hivon2002} to distinguish it from the alternative ``$C^{-1}$" method used for \WMAP{}9 (see Section 2.1).
 \item Calculate the sample covariance of the simulated spectra (hereafter the simulated covariance).
 \item To reduce the random fluctuations in the simulated covariance and the bias to its inverse matrix due to the finite number of simulations \citep{Sellentin:2015waz}, we apply the same binning scheme to the simulated spectra and covariance matrices (both the analytic and the simulated) as was applied in the published \planck{} 2015 likelihood code\footnote{\url{http://pla.esac.esa.int/pla/\#cosmology}}. The number of bins that cover $30\leq\ell\leq1200$ is 136.
 \item Calibrate the analytic covariance using the simulations, as described in Section 2.3. The calibrated matrix is referred to as the corrected analytic covariance, which we will be using for our final analysis. The analytic covariance matrix underestimates the true covariance by up to 10\% for some multipoles (see Appendix A), and the simulations are used to correct for this.

\item  With the binned corrected analytic covariance, we follow procedures described in Section 4 to derive the covariance of the difference between the observed \WMAP{}9 and \planck{} 2015 TT spectra and test whether this difference is consistent with zero.
\end{enumerate}

More details of the procedure are provided in the following subsections.
\begin{figure*}
\begin{center}
\includegraphics[width=18cm]{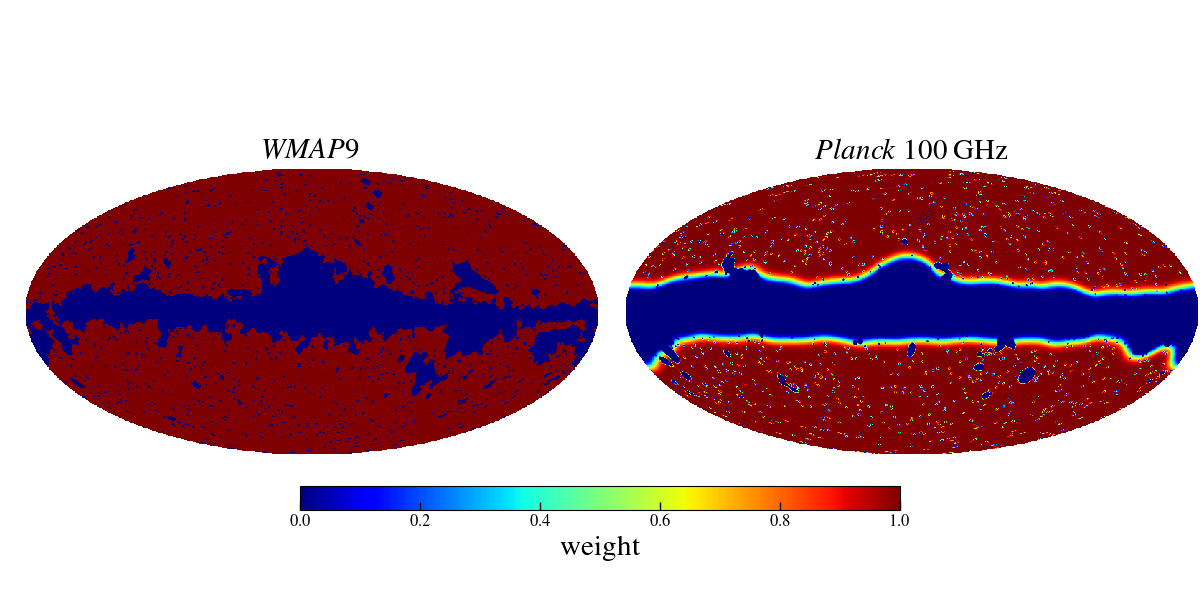}

\caption{Left: the KQ85y9 temperature analysis mask used in the \WMAP{}9 analysis \citep{Bennett2013}; right: the \planck{} 2015 T66 mask for 100 GHz \citep{planck/11:2015}. As shown, not all the areas chosen to be masked are the same. The unmasked fraction of the sky is 75\% for \WMAP{}9 and 66\% for \planck{} 100 GHz. While the \WMAP{}9 mask only has weights of 0 and 1, the \planck{} mask is apodized, with weights in between. The difference between \WMAP{}9 and \planck{} 2015 sky masks is the reason that the \WMAP{}9 and the \planck{} 2015 spectra are not fully correlated even at  $\ell\leq300$, where noise is negligible compared to cosmic variance uncertainty. See Section 2.3. \label{fig:mask}  }
\end{center}

\end{figure*}

\subsection{Simulating \WMAP{}9 spectra}
The \WMAP{} instrument was composed of 10 differencing assemblies (DAs) spanning five frequencies from 23 to 94 GHz \citep{Bennett2003a}. The three lowest frequency bands are used as foreground monitors. Only the V band ($\sim $ 61 GHz) and the W band ($\sim $ 94 GHz) are used to compute the TT spectrum \citep{wmap3Hinshaw}. The beam widths for the V and the W band are ${0.33}^{\circ}$ and ${0.21}^{\circ}$ FWHM respectively. In the \WMAP{} analysis, the $a_{\ell m}$ coefficients were computed from the \texttt{Healpix} $N_{\mathrm{side}}$ = 1024 (10 arcmin pixels) maps for each single year and each single-DA (V1, V2, W1-W4). For low multipoles ($2\leq\ell < 30$) a pixel-based likelihood was used, while a power spectrum based likelihood was used for $30\leq \ell \leq1200$. Until the nine-year release of \WMAP{} data, for $\ell \leq 600$ the coefficients were evaluated with uniform pixel weighting, which is optimal in the signal-dominated region, while inverse-noise weighting, optimal in the noise-dominated region, was used for $\ell > 600$ \citep{wmap7}. The TT cross-power spectra are computed from all the pairs of independent maps. For \WMAP{}9\footnote{For the \WMAP{}9 likelihood code, see \url{https://lambda.gsfc.nasa.gov/product/map/current/likelihood_get.cfm}.}, a different power spectrum estimator, the $C^{-1}$ method, was used \citep{Bennett2013}. However, it would be computationally challenging to implement $C^{-1}$ on all our 4000 simulations. We will show in Section 4 that our results should not be affected significantly by the change of $C_\ell$ estimator. The \WMAP{}9 analysis also took into account the uncertainties from the beam functions and point sources, but we exclude these because their effect is small (contributing about $0.06\%$ of the $\ell>30$ temperature log-likelihood) and is not expected to correlate with \planck{} uncertainties.

Using the best-fit power spectrum of \WMAP{}9 TT data with the reionization optical depth $\tau$ fixed to be 0.07 (see Table \ref{param} for the model parameters), we generate 4000 realizations. At $30\leq \ell \leq 1200$, $\tau$ is strongly degenerate with $A_s$, the amplitude of primordial density fluctuations, as  the TT spectrum is only sensitive to the parameter combination $A_s e^{-2\tau}$. Fixing $\tau$ breaks the degeneracy, and the value 0.07 is also consistent with those inferred by \WMAP{}9 and \planck{} 2015 data, being within $1.5\sigma$ of their constraints \citep[see Table 1 of][for recently published values of $\tau$ from different choices of data sets]{weiland2018}.

To simulate the maps observed by each DA, we multiply the simulated spherical harmonics with the \WMAP{}9 beam window function for that DA and add Gaussian noise. We make one noise map for each DA by inverting the sum of inverse variances from maps in different years.
Then we apply the KQ85y9 temperature analysis mask \citep{Bennett2013}, which masks both galactic emission and bright point sources, leaving 75\% of the sky to be analyzed, see Figure \ref{fig:mask}. Next we compute cross spectra for the six realistic maps with the appropriate pixel weighting applied to their corresponding range of multipoles.

\begin{figure*}
\begin{center}
\includegraphics[width=14cm]{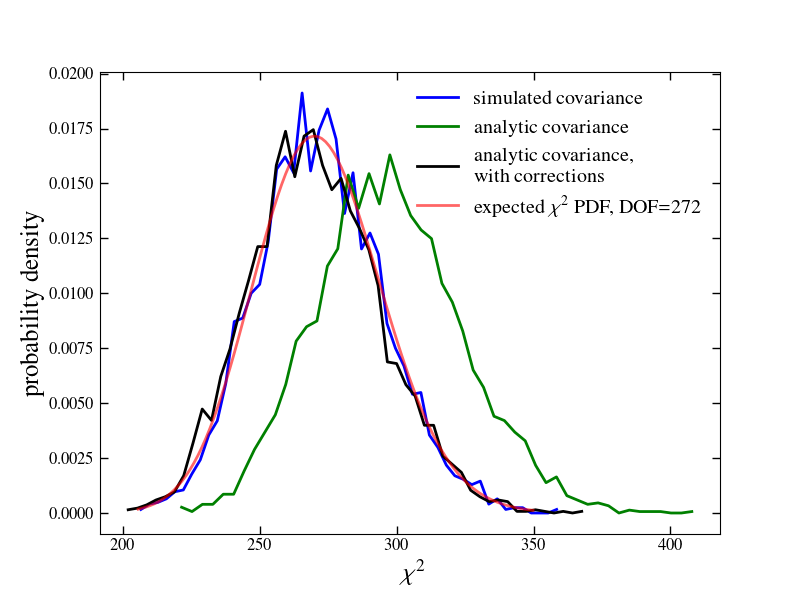}

\caption{The $\chi^2$ distribution of 4000 simulated binned spectra of \WMAP{}9 and \planck{} 2015, using different versions of binned covariance matrices. The analytic covariance (green) produces a $\chi^2$ higher than expected, consistent with the fact that it underestimates the true covariance of the simulated spectra (see text and Appendix A). The simulated covariance (blue) leads to a slightly narrower histogram, as the finite number of simulations introduces a slight bias into the inverse simulated covariance. The corrected analytic covariance (black) recovers the expected $\chi^2$ probability density function (PDF), with 272 degrees of freedom (DOF). This indicates that the binned power spectrum is well approximated by a Gaussian distribution with the mean and the covariance matching the fiducial spectrum and the corrected analytic covariance matrix, respectively. Thus, for the subsequent analysis, we use the corrected analytic covariance matrix. \label{fig:chi2_2x2} }
\end{center}
\end{figure*}

\subsection{Simulating \planck{} 2015 spectra}
The \planck{} instrument consists of seventy-four detectors in nine frequency bands between 30 and 857 GHz \citep{planck/01:2013}.  Similarly to \WMAP{}9, a pixel-based likelihood is used at $2\leq\ell<30$, and  a power spectrum based likelihood is used for $30\leq \ell \leq 2508$. For $\ell \geq 30$, TT spectra are computed as cross-spectra between the first half-mission and the second half-mission maps of different detector combinations, in three frequency channels: 100 GHz, 143 GHz and 217 GHz. Their effective beam FWHM in arcmin are 9.68, 7.30 and 5.02 respectively \citep{planck/11:2015}. Different masks are applied to the half-mission maps for each frequency. The masks applied are T66, T57 and T47 for 100 GHz, 143 GHz and 217 GHz, respectively. See Figure \ref{fig:mask} for the T66 mask of the 100 GHz temperature maps. The final power spectrum is an optimal combination of the 100$\times$100 GHz, 143$\times$143 GHz, 217$\times$217 GHz and 143$\times$217 GHz spectra.

The procedure of simulating \planck{} 2015 spectra is very similar to that of \WMAP{}9, except we make use of the published \planck{} 2015 half-mission noise maps and simulate six CMB signal maps for the three frequencies mentioned above and their two half-missions. We include the effect of the published beam window functions and work at $N_{\mathrm{side}} =$ 2048, corresponding to 5 arcmin pixels. We ignore the noise correlation between pixels \citep{planck/11:2015}, since the \planck{} and \WMAP{} noise are independent and only enter the \WMAP{}9 $\times$ \planck{} 2015 covariance matrix indirectly through the weighting of power spectra. Then we apply the masks for each frequency and obtain the cross spectra up to $\ell = 1200$ for the same four frequency combinations used in the experiment. 
We do not include \planck{} foregrounds in the simulations because the dominant Galactic and extragalactic dust foregrounds in the \planck{} channels are far smaller at the lower \WMAP{} frequencies. We therefore do not expect foreground uncertainties to contribute significantly to the \WMAP{}-\planck{} correlation.

\begin{figure*}
\begin{center}
\includegraphics[width=18cm]{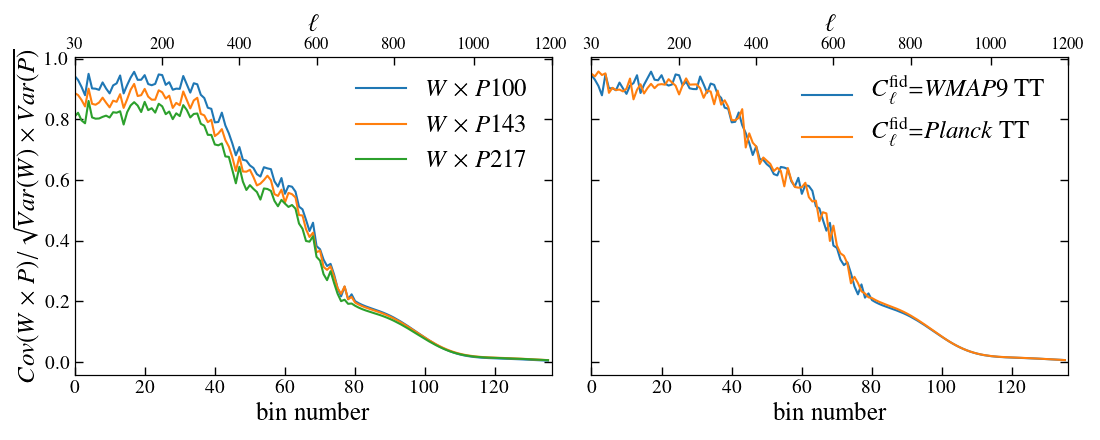}

\caption{ The correlation between \WMAP{}9 ($W$) and \planck{} 2015 ($P$) binned TT power spectra, defined as the ratio of the diagonal elements of the corrected analytic covariance between the combined spectra, to the square root of the product of the experimental variances. The axis on the top shows the center multipole of each bin. The spiky structure in the first 80 bins is due to calibrating the analytic covariance using the simulations, which introduces small random fluctuations. Left: Comparison of the correlation between \WMAP{}9 and different \planck{} 2015 frequency channels, with the \WMAP{}9 best-fit spectrum as the fiducial spectrum. The \WMAP{} mask uses 75\% of the sky while the sky fractions of the masks for \planck{} 100, 143, and 217 GHz are 66\%, 57\%, and 47\%, respectively. The correlation falls off at smaller scales as \WMAP{} variance becomes dominated by noise. \planck{} masks with lower sky fraction produce lower correlation with \WMAP{}9. Right: We also compare the correlation between the combined spectra using different fiducial models for simulations, the best-fit spectrum from \WMAP{}9 $\ell \geq 30$ data, and \planck{} 2015 $\ell \geq 30$. In Section 4, we show that the choice of fiducial spectra makes a negligible difference. \label{fig:correlation} }

\end{center}
\end{figure*}

\subsection{Calculating power spectrum covariance}
We follow the procedure in Appendix C.1.1 of \cite{planck/11:2015} applying the MASTER method \citep{myth} to derive a full, analytic covariance for both experiments accounting for the effect of noise, window functions, masks and different map weighting schemes. The main idea of the procedure is that first we calculate the power spectrum of a masked map, then we perform mask deconvolution to recover an unbiased estimate of the true underlying spectrum. Next we bin both the simulated spectra and analytic covariance matrices, using the binning matrix $B$ provided by the \planck{} 2015 likelihood code. The binned spectra and covariance are obtained from the following expressions:
 \begin{eqnarray}
 C_b &=&\sum_{\ell} B_{b\ell} C_\ell, \\
 \bold{\Sigma}_{XY,bb'}&=&\sum_{\ell,\ell'} B_{b\ell} \bold{\Sigma}_{XY,\ell\ell'} B^T_{\ell'b'}
\end{eqnarray}
where $b$ runs over 136 bins and $\bold{\Sigma}$ is a covariance matrix. The subscripts $X$ and $Y$ are either $W$ or $P$, referring to \WMAP{}9 and \planck{} 2015 respectively. The measured/simulated $C_\ell$s are only approximately $\chi^2$ distributed due to masking. With the large number of modes being combined into each bin, the $C_b$s can be well approximated as Gaussian \citep{planck/11:2015}.

We then co-add the spectra based on their inverse covariance to obtain one combined spectrum for \WMAP{}9 and one for \planck{} 2015, as well as the covariance matrices for the combined spectra, following the steps in Appendix C of \cite{wmap} and in Appendix C.4 of \cite{planck/11:2015} respectively.

As noted in \cite{planck/11:2015}, the analytic covariance, though not subject to random fluctuations in the simulations, does not fully capture the covariance of the simulated power spectra. We believe the disagreement arises from an assumption made in the analytic calculation that there is negligible variation over a small range of multipoles in the power spectrum. 
This leads to underestimation around $\sim 10\%$ for signal dominated regions (see Appendix A). To correct for such discrepancies, first we break down the covariance matrix  $\bold{\Sigma}$ into 4 sub-blocks as \begin{equation} \bold{\Sigma} = \left(
\begin{array}{cc}
 \bold{\Sigma}_{WW} & \bold{\Sigma}_{WP} \\
   \bold{\Sigma}_{PW} & \bold{\Sigma}_{PP} \\
  \end{array}
  \right)
\end{equation}
where each term is a 136$\times$136 matrix, and 136 is the number of bins for $30\leq \ell \leq 1200$. The elements $\bold{\Sigma}^\mathrm{ANA}_{XY,ij}$ in each sub-block of the analytic matrix are rescaled by the factor $r_{i}^2= \bold{\Sigma}^\mathrm{SIM}_{XY,ii}/ \bold{\Sigma}^\mathrm{ANA}_{XY,ii}$ which compares the simulated diagonal elements of one sub-block to the analytic. Then we rescale all the elements so that $\bold{\Sigma}^\mathrm{ANA,corrected}_{XY,ij}=\bold{\Sigma}^\mathrm{ANA}_{XY,ij}r_i r_j$. For the \WMAP{}-\planck{} analytic covariance, the correction is applied only to the first 80 of 136 bins. For bin numbers over 80, the scatter in the simulated covariance due to the \WMAP{} noise is much larger in magnitude than the analytic estimation.

Figure \ref{fig:chi2_2x2} shows the $\chi^2$ distribution of 4000 simulated, binned and combined spectra of \WMAP{}9 and \planck{} 2015, with 272 degrees of freedom. Here $\chi^2$  is defined as
\begin{equation} \chi^2= \sum^{272}_{b,b'=1} (\hat{C}^\mathrm{SIM}_{b}-C^\mathrm{fid}_b) ({\bold{\Sigma}}^{-1})_{bb'}(\hat{C}^\mathrm{SIM}_{b'}-C^\mathrm{fid}_{b'})
\end{equation}
where $C^\mathrm{fid}$ consists of two copies of the binned fiducial spectra and $\hat{C}^\mathrm{SIM} = (\hat{C}^\mathrm{SIM}_{W},\hat{C}^\mathrm{SIM}_{P})$ contains the simulated \WMAP{} and \planck{} spectra. The different lines in Figure \ref{fig:chi2_2x2} show results with different choices of $\bold{\Sigma}$: the simulated, the analytic, or the analytic with corrections. For the subsequent analysis, we use the corrected analytic covariance matrix.


We show in Figure \ref{fig:correlation} the correlation between \WMAP{}9 and \planck{} 2015 TT spectra, defined as the ratio of the diagonal elements of covariance between the two experiments, based on analytic calculation and calibrated by simulations, to the square root of the product of the experimental variances of \WMAP{}9 and \planck{} 2015. The correlation falls from 0.8-0.9 at low multipoles, where both experiments are cosmic variance limited, to close to zero at higher multipoles, where\WMAP{} variance is increasingly dominated by noise. The right panel of the figure shows that the correlation depends very weakly on the chosen fiducial spectrum.


\begin{figure}
\includegraphics[width=9cm]{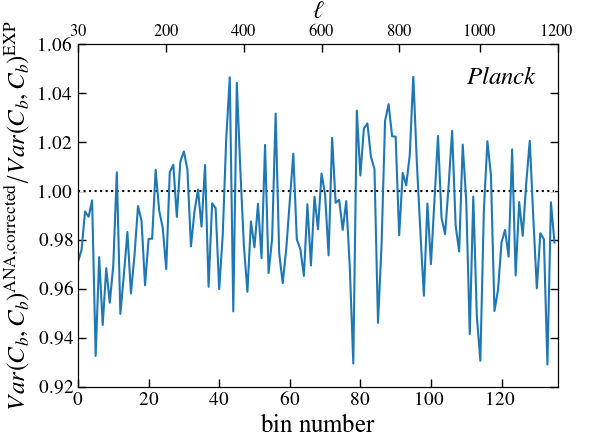}\\

\includegraphics[width=9cm]{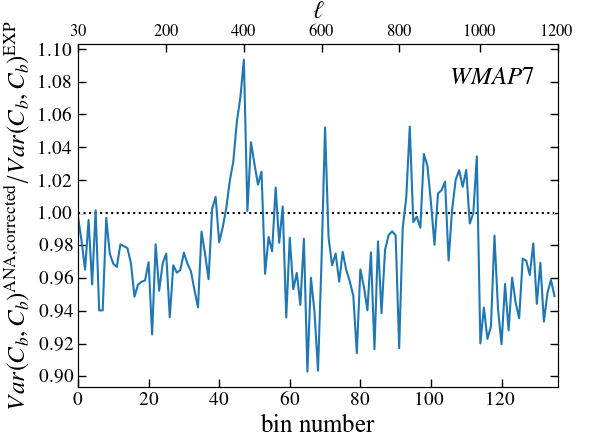}

\caption{Ratio of the corrected analytic binned TT variance to the experimental variance for \planck{} 2015 (top) and \WMAP{}7 (bottom).
We show the \WMAP{}7 ratio instead of \WMAP{}9 because both our simulations and \WMAP7{} use the MASTER power spectrum estimator, while \WMAP{}9 uses the $C^{-1}$ estimator. The deviations from unity are due to differences between our simulations and the analysis process of the experiments. The simulated \WMAP{}-\WMAP{} and \planck{}-\planck{} covariances are not used in our final consistency test. See text and Section 4 for discussion of implications for the \WMAP{}-\planck{} covariance. \label{fig:ratio_cm_wp7} }

\end{figure}

\section{comparing simulations to experiments}
To test whether our simulations capture experimental properties, we compare the corrected analytic variance of simulated power spectra to the ones published by the \WMAP{} and \planck{} teams. Since we are only using the simulations and analytic calculations to estimate the cross-covariance between \WMAP{} and \planck{}, the exact agreement for the $W\times W$ and the $P\times P$ covariance is not required. For \WMAP{}9, the variance provided by the published likelihood code depends on the choice of the theory spectrum. We choose to use the best-fit power spectrum of \WMAP{}9 TT data with a fixed $\tau$, so as to be consistent with our simulations. 
On the other hand, the \planck{} 2015 published variance is based on a fixed fiducial spectrum fit to $\ell\geq30$ with $\tau=0.07\pm0.02$ \citep[Section 3.3 of][]{planck/11:2015}. We therefore used our simulations generated with the \planck\ $\ell\geq30$ best-fit model with $\tau=0.07$ as input when comparing to the published variance. As mentioned earlier, the exact choice of $\tau$ is not important for $30\leq\ell\leq1200$.

For \planck{} 2015, the variance used as reference is a binned matrix obtained from co-adding the covariance of different cross spectra provided by the \planck{} 2015 released likelihood code, following procedures similar to co-adding the spectra, as mentioned in Section 2.3. The simulated/experimental (S/E) ratio, shown in the top panel of Figure \ref{fig:ratio_cm_wp7}, is on average slightly below 1. We believe this is due the fact that our simulations do not exactly replicate the \planck{} 2015 analysis process. We investigate the potential effect of this underestimation on the \WMAP{}-\planck{} covariance in Section 4.

For \WMAP{}9 the situation is more complicated.
In the analysis of the nine-year data, the \wmap\ team replaced the MASTER power spectrum estimator by an optimal $C^{-1}$ estimator. This estimator uses all the two-point correlation information in the unmasked pixels in the map, or, equivalently, the full covariance structure in the harmonics of the masked map, $\tilde{a}_{\ell m}$. Pseudo-$C_{\ell}$ methods like MASTER provide an unbiased estimate for the underlying power spectrum but only utilize products of $\tilde{a}_{\ell m}$ for the same $\ell$ and $m$ (see Appendix A), causing some loss of information \citep[e.g.,][]{gruetjen/shellard:2014}. The published \WMAP{}9 likelihood package does not include results analyzed using the MASTER method, so we generate another set of 4000 simulations with \WMAP{} seven-year (\WMAP{}7) data properties and compare their spectrum variance to the result from inverting the Fisher matrix in the \WMAP{}7 likelihood code. The bottom panel of Figure \ref{fig:ratio_cm_wp7} shows the S/E ratio for \WMAP{}7. 
Numerical differences exist between our analytic calculation using the MASTER method and the approximation used for the \WMAP{}7 Fisher matrix, causing deviations from unity in the S/E ratio. This difference is unlikely to have any significant effect on our final results, because it is smaller than the difference between using the MASTER method and using the $C^{-1}$ method and even that does not change our conclusion, as discussed below.


Going from MASTER method to $C^{-1}$ reduces the power spectrum variance by 7-17\% as shown in Figure 31 of \cite{Bennett2013}.
This means our simulations with MASTER overestimate the experimental variance of \wmap9. Fortunately, this should not impact the \wmap-\planck\ covariance, which is what we are using the simulations to obtain, because the \planck\ analysis used MASTER. The additional information about the $C_{\ell}$s gained from applying the $C^{-1}$ estimator to \wmap\ maps is therefore not present in the \planck\ 2015 power spectra and should not lead to a reduction of the \wmap-\planck\ covariance.

In Section~4 we test this argument by investigating the effect of different pixel weightings of the \wmap9 temperature maps on the \wmap-\planck\ covariance. The different weighting schemes represent more extreme changes in  the \wmap9 TT uncertainties than changing from MASTER to $C^{-1}$, but do not lead to changes to our conclusion about the consistency of the experiments.

\begin{table}\small
\setlength\tabcolsep{6pt}

\begin{center}
\renewcommand{\arraystretch}{1.5}
\begin{tabular}{cccccccccc}
\\
\hline\hline
$C^\mathrm{fid}_\ell$  & \WMAP{}9 Pixel Weighting & $\chi^2_\mathrm{diff} $ & PTE \\ \hline

\WMAP{}9      & Hybrid
                   & 141.8  & 0.35 

                                     \\
\planck{} 2015 & Hybrid & 139.6 &   0.40 

                   \\

\WMAP{}9       & Uniform
                   & 150.7 & 0.18 

                   \\

\WMAP{}9                & Inv Noise
                   & 139.4 &0.40 

\\
\hline
\end{tabular}
\end{center}
\caption{$\chi^2_\mathrm{diff}$ and PTE results for the observed power spectrum difference, with different fiducial input power spectra for simulations, and different weighting schemes for \WMAP{}9 maps. The degree of freedom is 136. Three different weighting schemes are applied to the simulated \WMAP{}9 temperature maps. Uniform is when all the pixels share the same weight. Inverse noise weighting weights the pixels by their inverse noise variance. Hybrid is using uniform weighting for $\ell \leq 600$ and inverse noise for $\ell > 600$. We find no significant difference in the values of $\chi^2_{\rm{diff}}$ and PTE, using different fiducial spectra or different weighting schemes. We conclude that there is no significant difference between the observed \WMAP{}9 and \planck{} 2015 TT spectra over their common multipole range. \label{chi2_PTE}   }
\end{table}

\begin{figure*}
\begin{center}
\includegraphics[width=18cm]{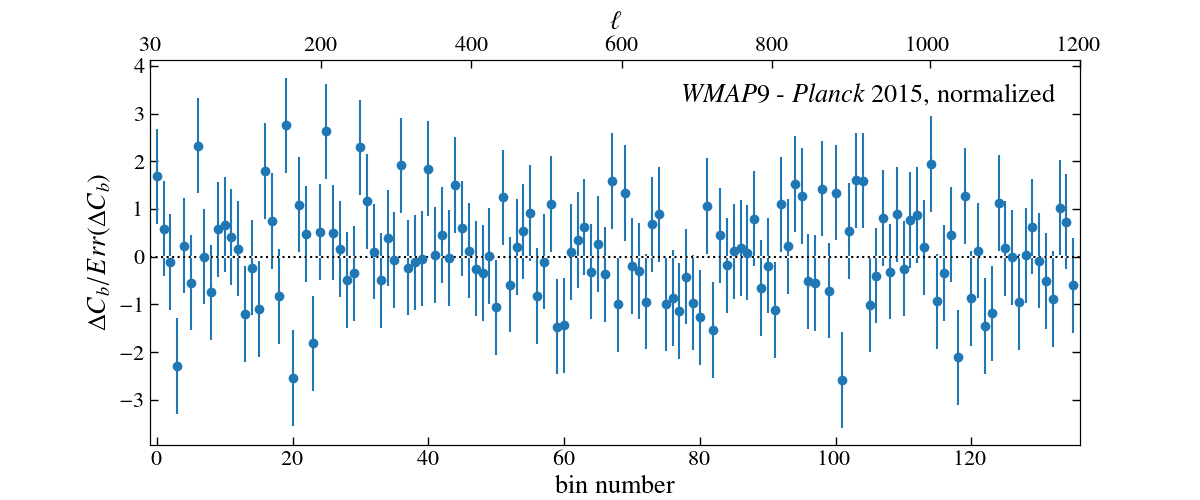}

\includegraphics[width=18cm]{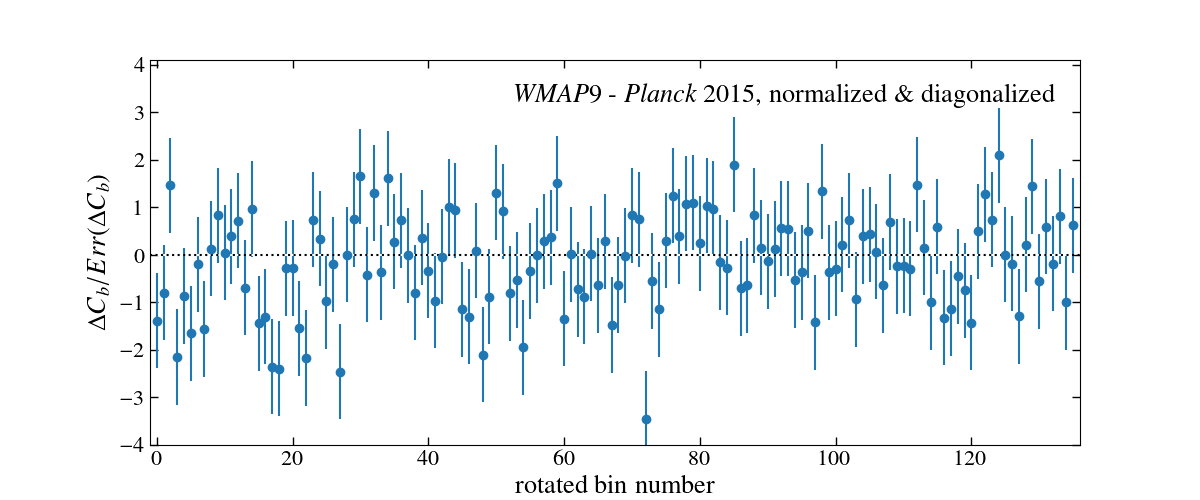}

\caption{Top: observed binned power spectrum difference between \WMAP{}9 and \planck{} 2015, normalized by error bars estimated from simulations, which account for the correlated CMB cosmic variance between the two experiments. Most data points are within 2$\sigma$ from zero. The first 13 bins are anti-correlated at $\sim$ 13\% with their immediate neighbors, while the rest are at $\sim$ 5\%. Bottom: the vector of differences is rotated so that its covariance is diagonalized and the bins are uncorrelated. The rotated difference shows no statistically significant deviation from zero, except for the 72nd bin. We do not consider it as a sign of inconsistency, because the probability of at least 1 out of 136 bins deviating more than $3\sigma$ from zero is 25\%, for 136 independent Gaussian-distributed random variables. 
We note that similar ``clumping" of adjacent points also appears in randomly generated sets of 136 Gaussian numbers. \label{fig:cl_diff}  }
\end{center}

\end{figure*}

\section{quantifying consistency}
To compare results from \WMAP{}9 and \planck{} 2015, we need the power spectrum difference array $\Delta C_b$ and its associated covariance $\Delta \bold{\Sigma}$. The latter is given by
\begin{equation}
\Delta \bold{\Sigma} =\bold{\Sigma}_{WW} + \bold{\Sigma}_{PP} - \bold{\Sigma}_{WP}  - \bold{\Sigma}_{PW}
\end{equation}
and $\Delta C_b = C^{\mathrm{OBS}}_{W,b} - C^{\mathrm{OBS}}_{P,b} $ is the observed difference of binned power spectra in the common range of $\ell$, provided by the two experiments. Then we calculate the $\chi^2$ of the difference defined by
\begin{equation}\chi^2_\mathrm{diff}=\sum^{136}_{b,b'=1}\Delta C^T_b \Delta \bold{\Sigma}^{-1}_{bb'} \Delta C_b\end{equation}
and its probability to exceed (PTE) for a $\chi^2$ distribution with 136 degrees of freedom (the number of bins). Finally we convert the PTE values to an equivalent number of Gaussian standard deviations. 

For $\bold{\Sigma}_{PP}$, we bin and co-add the covariance matrices for the 4 frequency combinations provided by the \planck{} 2015 likelihood code while $\bold{\Sigma}_{WW}$ is from inverting the Fisher matrix calculated from the \WMAP{}9 likelihood code. For $\bold{\Sigma}_{WP}$  and $ \bold{\Sigma}_{PW}$ we use the corrected analytic $W\times P$ and $P\times W$ covariance matrices described in Section~2.3.

The $\chi^2_\mathrm{diff}$ and PTE of the observed power spectrum difference are shown in Table \ref{chi2_PTE}. Using different input fiducial spectra or different pixel weighting schemes on simulated \WMAP{}9 temperature maps does not change the values of $\chi^2_\mathrm{diff}$ or PTE significantly. The cases closest to the actual experiments are the ones using hybrid weighting for simulated \WMAP{}9 maps. Using the \WMAP{}9 best-fit TT spectrum as the fiducial gives PTE 0.35, which means the \planck{} 2015 observed TT spectrum differs from \WMAP{}9 at only 0.39$\sigma$, while using \planck{} 2015 best-fit spectrum as fiducial gives PTE 0.40, corresponding to a 0.26$\sigma$ difference. This leads to the conclusion that there is no significant difference between the observed \WMAP{}9 and \planck{} 2015 TT spectra over their common multipole range ($30\leq\ell \leq1200$), regardless of the choice of assumed models. Choosing a fiducial model with more drastically different cosmological parameters could have a larger impact on the \wmap-\planck\ consistency, however there is no motivation to consider such a model as it would provide a poor fit to the actual \wmap\ or \planck\ TT measurements.

Using different weightings on simulated \WMAP{}9 maps does not change our conclusion. We test the extreme case of comparing results between using uniform weighting for all $\ell$ and using inverse noise weighting for all $\ell$. The PTE value shifts from 0.18 to 0.40, corresponding to $0.91\sigma$ and $0.25\sigma$ respectively. The stability of our results against the change of map weighting schemes implies that our conclusion about the consistency between \WMAP{}9 and \planck{} 2015 TT spectra would not change even if we were to generate simulations using the $C^{-1}$ pipeline as the \WMAP{} team did.

We test that the slight underestimation of experimental variances shown in Figure \ref{fig:ratio_cm_wp7} would make no significant difference to our results, even if this also affects the \WMAP{}-\planck{} covariance. Using the \WMAP{}9 best-fit TT spectrum as the fiducial input and hybrid pixel weightings on simulated \WMAP{} maps, rescaling the \WMAP{}-\planck{} covariance by a factor of $(1.011\times 1.024)^{0.5}$, which approximately compensates the underestimation, produces $\chi^2_\mathrm{diff} = 144.3$ and PTE 0.30, corresponding to a $0.53\sigma$ difference.

The top of Figure \ref{fig:cl_diff} illustrates the observed power spectrum differences of binned power spectrum $\Delta C_b$, with error bars given by $\Delta \bold{\Sigma}$ accounting for the correlated cosmic variance between \WMAP{}9 and \planck{} 2015. To facilitate visual comparison, we divide the differences by their uncertainties, so that all the error bars are unity. As shown in this figure, the observed difference is roughly consistent with zero. Small correlations between adjacent bins ($\sim -13\%$ for the first 13 bins and $\sim -5\%$ for the rest) are accounted for when calculating $\chi_{\rm diff}^2$ but make visual assessment of $\Delta C_b$ more difficult. We therefore apply a rotation to the vector of differences so that its covariance is diagonalized and the bins are uncorrelated. The rotation matrix $\bold{U}$ is constructed from the eigenvectors of $\Delta \bold{\Sigma}$. The rotated vector of difference $\Delta C^R$ and its covariance $\Delta \bold{\Sigma}^R$ are given by the following:
\begin{eqnarray}
\Delta C^R &=& \bold{U}^{-1} \Delta C \\
\Delta \bold{\Sigma}^R &=& \bold{U}^{-1} \Delta \bold{\Sigma}\bold{U}
\end{eqnarray}

The resulting normalized difference is shown in the bottom of Figure \ref{fig:cl_diff}, showing no statistically significant deviation from zero, except for the 72nd bin. Assuming all the uncorrelated bins are Gaussian distributed, the probability of at least 1 out of 136 bins deviating more than $3\sigma$ from zero is 25\%. Therefore we do not take this as a sign of inconsistency. Moreover, we do not think the apparent ``clumping'' of data points is anything more than statistical fluctuations. Human eyes are naturally drawn to patterns and thus tend to discover ``features". 
An example test for the occurrence of clumping is to ask whether the maximum number of consecutive points lying above or below zero (8 for the bottom panel of Figure \ref{fig:cl_diff}, see the 16th to 23rd bins) is unusually high. We generated 10000 sets of 136 independent normally distributed values and found that 23.1\% included 8 or more consecutive points lying above or below zero, indicating that the behavior in Figure \ref{fig:cl_diff} is consistent with statistical fluctuations.

In addition to the full $30\leq\ell\leq1200$ range we also calculated the $\chi^2$ and PTE from different subsets of multipoles, including varying the maximum multipole. The PTE values from these tests were largely between 0.05 and 0.4, however we found that restricting the comparison to, for example, $30\leq\ell\leq200$ (up to bin 26), or $30\leq\ell\leq300$ (bin 37), produced lower PTE values of 0.005 and 0.012. Of our 4000 simulated \wmap\ and \planck\ spectra, 400 (10.0\%) produce PTE values less than $0.01$ as the maximum multipole was varied, and 491 (12.3\%) produced values greater than $0.99$. Restricting to the 373 realizations with PTE between 0.3 and 0.4 for $30\leq\ell\leq1200$, similar to the data value of 0.35, the corresponding numbers are 26 (7.0\%) for PTE values less than $0.01$ and 22 (5.9\%) for greater than $0.99$. The data therefore do not appear particularly anomalous in this respect.

\begin{figure*}
\begin{center}
\includegraphics[width=15cm]{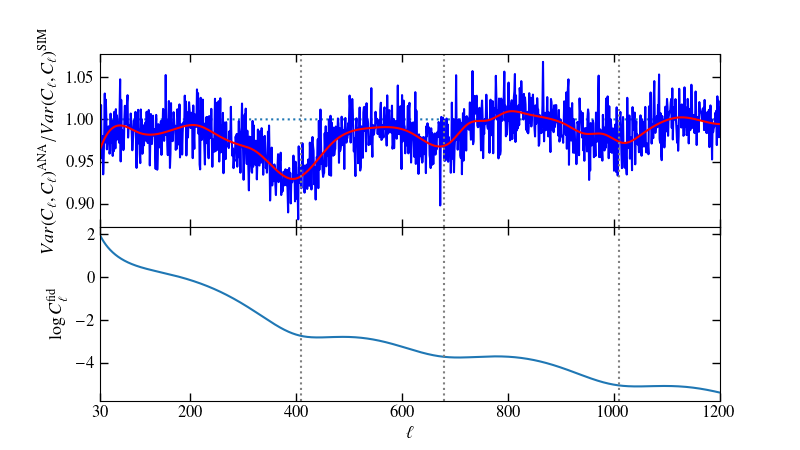}

\caption{ Top: ratio of the analytic variance to the simulated, from noise free simulations. The blue is the raw ratio and the red line is a smooth fit based on cubic splines. Bottom: the logarithm of the fiducial spectrum. The vertical dotted lines show that the minima in the ratio correspond to the multipoles where the drop of the power spectrum reaches a temporary plateau. These are the multipoles where the approximation that the spectrum does not vary over a small range of neighboring $\ell$s is most unrealistic.   \label{fig:ana_sim_cm_ratio}  }
\end{center}

\end{figure*}

\section{conclusions}
We quantify the consistency between the observed TT power spectra from \WMAP{}9 and \planck{} 2015 over their overlapping multipole range where power spectrum based likelihoods were used. We generated simulations to account for the cosmic variance common to both experiments.
Their correlation is estimated to be as high as $\sim 90\%$ in signal-dominated regions (roughly $\ell < 300$), and drops below 10\% roughly at $\ell>850$. Even taking into account their correlation, we find that the spectra are consistent within $1\sigma$. We also note that with the common cosmic variance taken out of the covariance of the power spectrum difference between the two experiments, the consistency test presented in Section 4 is more sensitive to any unknown systematics or underestimated \WMAP{} noise, than any test that can be done with each experiment alone.

We also tested our simulation fidelity in Section 2 and 3. We find that our simulated power spectra are consistently Gaussian distributed, with the mean being the input fiducial spectrum and the covariance properly estimated.



While we did not implement the optimal $C^{-1}$ estimator on simulated \WMAP{} maps as in the \WMAP{}9 analysis, we tested the impact of pixel weighting on the \WMAP{}-\planck{} covariance from adopting two extreme weighting schemes.  We found that using either uniform weighting at all multipoles, or inverse-noise weighting at all multipoles, still resulted in agreement with \planck{} within $0.91\sigma$. The different weightings mainly affect the \WMAP{} noise contribution, which does not enter into the \WMAP{}-\planck{} covariance. We also demonstrated the stability of our results against the choice of fiducial spectrum used in the simulations. Using the best-fit spectrum of the \planck{} 2015 TT data instead of that of \WMAP{}9 only impacts the comparison at around $0.1\sigma$.

The consistency shown in our analysis provides high confidence in both the \WMAP{}9 temperature power spectrum and the overlapping multipole region of the \planck{} 2015 power spectrum, virtually independent of any assumed cosmological model. The \planck{} 2018 TT spectrum is only minimally different from the 2015 version \citep{planck2018VI}, and we therefore expect it to remain consistent with \wmap.

The difference between cosmological constraints from \WMAP{} and \planck{} TT spectra is driven by higher multipoles in \planck{}, which also drive the tensions with some astrophysical data discussed earlier. An important check of these \planck{} measurements will come from similar tests to those performed in this work using temperature and polarization measurements from high-resolution experiments \citep[e.g.,][]{louis/etal:2014,hou/etal:2018}.\\


This research was supported in part by NASA grants NNX14AF64G, NNX16AF28G, and NNX17AF34G. We acknowledge the use of the Legacy Archive for Microwave Background Data  Analysis  (LAMBDA),  part  of  the  High  Energy Astrophysics Science Archive Center (HEASARC). HEASARC/LAMBDA is a service of the Astrophysics Science Division at the NASA Goddard Space Flight Center. This work was partly based on observations obtained with Planck (http://www.esa.int/Planck), an ESA science mission with instruments and contributions directly funded by ESA Member States, NASA, and Canada. Part of this research project was conducted using computational resources at the Maryland Advanced Research Computing Center (MARCC).

The authors would like to thanks Keisuke Osumi and Duncan Watts for reading a draft of the manuscript and providing comments and suggestions

\appendix
\section{Underestimation of variance due to assumptions in analytic calculations}

In this appendix, we discuss why the covariance obtained from analytic calculations based on the MASTER method underestimates the simulated one. This is also noted in Appendix C.1.4 of \cite{planck/11:2015}. The upper panel of Figure \ref{fig:ana_sim_cm_ratio} shows an example of the ratio of the analytic to the simulated variance, calculated from noise free simulations. 

Limiting ourselves to the noiseless case, we look at a few key equations in the MASTER method and point out when the approximation is made and how it affects the result of the calculation. For a detailed description of the method, see \cite{myth}.

The spherical harmonic transform of a temperature map $\Delta T_i$ with mask $w_i$ is defined as \begin{equation} \tilde a_{\ell m} = \sum_i \Delta T_i w_i
\Omega_i Y_{\ell m}({\bf \theta}_i)\nonumber \end{equation} where $\Omega_i$ is the area of pixel $i$. The pseudo-harmonic $\tilde a_{\ell m}$s are related to the true $a_{\ell m}$s on the unmasked sky via a coupling matrix $K$:

\begin{equation}
 \tilde a_{\ell m} = \sum_{\ell^\prime m^\prime} a_{\ell^\prime m^\prime}
K_{{\ell m}{\ell^\prime m^\prime}}. \nonumber \end{equation} The detailed expression of $K$ is not important for this discussion, but we note that $K$ is close to 1 when $\ell = \ell' $, then drops off as $\ell'$ shifts away from $\ell$.

The pseudo-$C_\ell$ estimator is constructed from the sum

\begin{equation}
   \tilde C^p_\ell = {1 \over (2 \ell + 1)} \sum_m \vert \tilde a_{\ell m} \vert ^2\nonumber
\end{equation}
and its covariance is given by
\begin{eqnarray}
 \langle \Delta \tilde C^p_\ell
\Delta \tilde C^p_{\ell^\prime} \rangle &=&
 {2 \over (2 \ell + 1) (2 \ell^\prime + 1) }
\sum_{m m^\prime} \sum_{\ell_1 m_1} \sum_{\ell_2 m_2}
C_{\ell_1} C_{\ell_2}  \nonumber \\
&\times&  K_{\ell m \ell_1 m_1} K^*_{\ell^\prime m^\prime \ell_1 m_1} K^*_{\ell m \ell_2 m_2} K_{\ell^\prime m^\prime \ell_2 m_2}. \nonumber \end{eqnarray}
To make the calculation above computationally feasible, the power spectrum is approximated as unchanged over small range of multipoles where $\Delta \ell $ is small and $K$ is not negligible. Then $C_{\ell_1}$ and $C_{\ell_2}$ can be taken out of the summation and replaced by $C_\ell$ and $C_\ell'$. This will simplify the expression to the following:
\begin{equation}
 \langle \Delta \tilde C^p_\ell
\Delta \tilde C^p_{\ell^\prime} \rangle =  \tilde V_{\ell \ell^\prime}
\approx  2 C_\ell C_{\ell^\prime}
\Xi (\ell, \ell^\prime, \tilde W^{(2)})  \nonumber
\end{equation}
where \begin{equation}
  \Xi(\ell_1, \ell_2, \tilde W^{(2)})  =
\sum_{\ell_3 }  {(2 \ell_3 + 1) \over 4 \pi}\tilde W^{(2)} _{\ell_3}
{\left ( \begin{array}{ccc}
        \ell_1 & \ell_2 & \ell_3  \\
        0  & 0 & 0
       \end{array} \right )^2} \nonumber
\end{equation}
and $\tilde W^{(2)}_{\ell}$ is the power spectrum of the square of the mask $w_i$,
\begin{equation}
 \tilde W^{(2)}_\ell = {1 \over (2 \ell + 1)} \sum_m \vert \tilde
w^{(2)}_{\ell m} \vert^2, \qquad \tilde w^{(2)}_{\ell m} = \sum_i w^2_i \Omega_i Y_{\ell m} (\theta_i). \label{V2b}  \nonumber
\end{equation}

The effect of this approximation on the covariance is minimal where the power spectrum, $C_\ell$ is declining or about to. Restricting to the simpler case where $\ell=\ell'$, the reasoning is as follows.

When the spectrum is declining around a certain $\ell$, using $C_\ell$ to replace its neighboring multipoles means that the calculation is underestimating the contribution to the variance from $\ell_1,\ell_2 < \ell$ and overestimating from  $\ell_1,\ell_2 > \ell$. So on average, the effect partly evens out.

At the high-$\ell$ end of plateaus, the central $C_\ell$ is about the same as the neighboring ones at smaller multipoles and larger than those at larger multipoles. The approximation means that we treat the neighboring $C_{\ell_1}$ and $C_{\ell_2}$ at $\ell_1,\ell_2 > \ell$ to be as large as $C_\ell$. With the neighboring $K$s being small, we are overestimating the contributions from terms that are relatively small compared to the ``correct answer". The fractional difference is negligible.

However, when the spectrum is at the end of a slope and the start of a plateau, we use a central $C_\ell$ to approximate neighboring, larger $C_{\ell_1}$ and $C_{\ell_2}$ at $\ell_1,\ell_2 < \ell$. Unlike the case at the high-$\ell$ end of plateaus, we are now underestimating terms that are not so small compared to the ``correct answer", even with the neighboring $K$s being small. The fractional difference is more significant.

Figure \ref{fig:ana_sim_cm_ratio} demonstrates the correspondence between the troughs of the ratio of the analytic variance to the simulated and the locations in the $\log C_\ell$ where the drop of power spectrum reaches a temporary plateau, supporting our argument. We also note that at around $\ell =400$ where the drop is sharpest, the disagreement between analytic and simulated variance is largest.

For a flat noise spectrum, the above approximation is exactly valid. Therefore the analytic covariance is more accurate for simulated maps with white noise, particularly at higher multipoles where the spectrum is noise-dominated.

This deviation of analytic covariance from the simulated covariance is why we make corrections on the former based on the latter.

\bibliographystyle{apj}

\begin{thebibliography}{}
\expandafter\ifx\csname natexlab\endcsname\relax\def\natexlab#1{#1}\fi

\bibitem[{{Addison} {et~al.}(2016){Addison}, {Huang}, {Watts}, {Bennett},
  {Halpern}, {Hinshaw}, \& {Weiland}}]{addison2016}
{Addison}, G.~E., {Huang}, Y., {Watts}, D.~J., {et~al.} 2016, \apj, 818, 132

\bibitem[{{Bennett} {et~al.}(2003){Bennett}, {Halpern}, {Hinshaw}, {Jarosik},
  {Kogut}, {Limon}, {Meyer}, {Page}, {Spergel}, {Tucker}, {Wollack}, {Wright},
  {Barnes}, {Greason}, {Hill}, {Komatsu}, {Nolta}, {Odegard}, {Peiris},
  {Verde}, \& {Weiland}}]{Bennett2003a}
{Bennett}, C.~L., {Halpern}, M., {Hinshaw}, G., {et~al.} 2003, \apjs, 148, 1

\bibitem[{{Bennett} {et~al.}(2013){Bennett}, {Larson}, {Weiland}, {Jarosik},
  {Hinshaw}, {Odegard}, {Smith}, {Hill}, {Gold}, {Halpern}, {Komatsu}, {Nolta},
  {Page}, {Spergel}, {Wollack}, {Dunkley}, {Kogut}, {Limon}, {Meyer}, {Tucker},
  \& {Wright}}]{Bennett2013}
{Bennett}, C.~L., {Larson}, D., {Weiland}, J.~L., {et~al.} 2013, \apjs, 208, 20

\bibitem[{{Chon} {et~al.}(2004){Chon}, {Challinor}, {Prunet}, {Hivon}, \&
  {Szapudi}}]{PolSpice2}
{Chon}, G., {Challinor}, A., {Prunet}, S., {Hivon}, E., \& {Szapudi}, I. 2004,
  \mnras, 350, 914

\bibitem[{{Efstathiou}(2004)}]{myth}
{Efstathiou}, G. 2004, \mnras, 349, 603

\bibitem[{{Fendt} \& {Wandelt}(2007)}]{pico}
{Fendt}, W.~A., \& {Wandelt}, B.~D. 2007, \apj, 654, 2

\bibitem[{{G{\'o}rski} {et~al.}(2005){G{\'o}rski}, {Hivon}, {Banday},
  {Wandelt}, {Hansen}, {Reinecke}, \& {Bartelmann}}]{Healpix}
{G{\'o}rski}, K.~M., {Hivon}, E., {Banday}, A.~J., {et~al.} 2005, \apj, 622,
  759

\bibitem[{{Gruetjen} \& {Shellard}(2014)}]{gruetjen/shellard:2014}
{Gruetjen}, H.~F., \& {Shellard}, E.~P.~S. 2014, \prd, 89, 063008

\bibitem[{{Hinshaw} {et~al.}(2003){Hinshaw}, {Spergel}, {Verde}, {Hill},
  {Meyer}, {Barnes}, {Bennett}, {Halpern}, {Jarosik}, {Kogut}, {Komatsu},
  {Limon}, {Page}, {Tucker}, {Weiland}, {Wollack}, \& {Wright}}]{wmap}
{Hinshaw}, G., {Spergel}, D.~N., {Verde}, L., {et~al.} 2003, \apjs, 148, 135

\bibitem[{{Hinshaw} {et~al.}(2007){Hinshaw}, {Nolta}, {Bennett}, {Bean},
  {Dor{\'e}}, {Greason}, {Halpern}, {Hill}, {Jarosik}, {Kogut}, {Komatsu},
  {Limon}, {Odegard}, {Meyer}, {Page}, {Peiris}, {Spergel}, {Tucker}, {Verde},
  {Weiland}, {Wollack}, \& {Wright}}]{wmap3Hinshaw}
{Hinshaw}, G., {Nolta}, M.~R., {Bennett}, C.~L., {et~al.} 2007, \apjs, 170, 288

\bibitem[{{Hivon} {et~al.}(2002){Hivon}, {G{\'o}rski}, {Netterfield}, {Crill},
  {Prunet}, \& {Hansen}}]{hivon2002}
{Hivon}, E., {G{\'o}rski}, K.~M., {Netterfield}, C.~B., {et~al.} 2002, \apj,
  567, 2

\bibitem[{{Hojjati} {et~al.}(2015){Hojjati}, {McCarthy}, {Harnois-Deraps},
  {Ma}, {Van Waerbeke}, {Hinshaw}, \& {Le Brun}}]{Hojiati2015}
{Hojjati}, A., {McCarthy}, I.~G., {Harnois-Deraps}, J., {et~al.} 2015, \jcap,
  10, 047

\bibitem[{{Hou} {et~al.}(2018){Hou}, {Aylor}, {Benson}, {Bleem}, {Carlstrom},
  {Chang}, {Cho}, {Chown}, {Crawford}, {Crites}, {de Haan}, {Dobbs}, {Everett},
  {Follin}, {George}, {Halverson}, {Harrington}, {Holder}, {Holzapfel},
  {Hrubes}, {Keisler}, {Knox}, {Lee}, {Leitch}, {Luong-Van}, {Marrone},
  {McMahon}, {Meyer}, {Millea}, {Mocanu}, {Mohr}, {Natoli}, {Omori}, {Padin},
  {Pryke}, {Reichardt}, {Ruhl}, {Sayre}, {Schaffer}, {Shirokoff},
  {Staniszewski}, {Stark}, {Story}, {Vanderlinde}, {Vieira}, \&
  {Williamson}}]{hou/etal:2018}
{Hou}, Z., {Aylor}, K., {Benson}, B.~A., {et~al.} 2018, \apj, 853, 3

\bibitem[{{Joudaki} {et~al.}(2018){Joudaki}, {Blake}, {Johnson}, {Amon},
  {Asgari}, {Choi}, {Erben}, {Glazebrook}, {Harnois-D{\'e}raps}, {Heymans},
  {Hildebrandt}, {Hoekstra}, {Klaes}, {Kuijken}, {Lidman}, {Mead}, {Miller},
  {Parkinson}, {Poole}, {Schneider}, {Viola}, \& {Wolf}}]{Joudaki2017}
{Joudaki}, S., {Blake}, C., {Johnson}, A., {et~al.} 2018, \mnras, 474, 4894

\bibitem[{{K{\"o}hlinger} {et~al.}(2017){K{\"o}hlinger}, {Viola}, {Joachimi},
  {Hoekstra}, {van Uitert}, {Hildebrandt}, {Choi}, {Erben}, {Heymans},
  {Joudaki}, {Klaes}, {Kuijken}, {Merten}, {Miller}, {Schneider}, \&
  {Valentijn}}]{Kohlinger2017}
{K{\"o}hlinger}, F., {Viola}, M., {Joachimi}, B., {et~al.} 2017, \mnras, 471,
  4412

\bibitem[{{Larson} {et~al.}(2015){Larson}, {Weiland}, {Hinshaw}, \&
  {Bennett}}]{larson2015}
{Larson}, D., {Weiland}, J.~L., {Hinshaw}, G., \& {Bennett}, C.~L. 2015, \apj,
  801, 9

\bibitem[{{Larson} {et~al.}(2011){Larson}, {Dunkley}, {Hinshaw}, {Komatsu},
  {Nolta}, {Bennett}, {Gold}, {Halpern}, {Hill}, {Jarosik}, {Kogut}, {Limon},
  {Meyer}, {Odegard}, {Page}, {Smith}, {Spergel}, {Tucker}, {Weiland},
  {Wollack}, \& {Wright}}]{wmap7}
{Larson}, D., {Dunkley}, J., {Hinshaw}, G., {et~al.} 2011, \apjs, 192, 16

\bibitem[{{Lewis} \& {Bridle}(2002)}]{cosmomc}
{Lewis}, A., \& {Bridle}, S. 2002, \prd, 66, 103511

\bibitem[{{Louis} {et~al.}(2014){Louis}, {Addison}, {Hasselfield}, {Bond},
  {Calabrese}, {Das}, {Devlin}, {Dunkley}, {D{\"u}nner}, {Gralla}, {Hajian},
  {Hincks}, {Hlozek}, {Huffenberger}, {Infante}, {Kosowsky}, {Marriage},
  {Moodley}, {N{\ae}ss}, {Niemack}, {Nolta}, {Page}, {Partridge}, {Sehgal},
  {Sievers}, {Spergel}, {Staggs}, {Walter}, \& {Wollack}}]{louis/etal:2014}
{Louis}, T., {Addison}, G.~E., {Hasselfield}, M., {et~al.} 2014, \jcap, 7, 16

\bibitem[{{McCarthy} {et~al.}(2018){McCarthy}, {Bird}, {Schaye},
  {Harnois-Deraps}, {Font}, \& {van Waerbeke}}]{McCarthy2018}
{McCarthy}, I.~G., {Bird}, S., {Schaye}, J., {et~al.} 2018, \mnras, 476, 2999

\bibitem[{{Planck Collaboration I}(2014)}]{planck/01:2013}
{Planck Collaboration I}. 2014, \aap, 571, A1

\bibitem[{{Planck Collaboration I}(2016)}]{planck/01:2015}
---. 2016, \aap, 594, A1

\bibitem[{{Planck Collaboration I}(2018)}]{planck2018I}
---.  2018, ArXiv e-prints, arXiv:1807.06205

\bibitem[{{Planck Collaboration VI}(2018)}]{planck2018VI}
{Planck Collaboration VI}. 2018, ArXiv e-prints, arXiv:1807.06209


\bibitem[{{Planck Collaboration VIII}(2016)}]{planck/8:2016}
{Planck Collaboration VIII}. 2016, \aap, 594, A8 
\bibitem[{{Planck Collaboration Int. XLVI}(2016)}]{planck/intermediate/46:2016}
{Planck Collaboration Int. XLVI}. 2016, \aap, 596, A107

\bibitem[{{Planck Collaboration Int. LI}(2017)}]{planck/intermediate/08:2016}
{Planck Collaboration Int. LI}. 2017, \aap, 607, A95

\bibitem[{{Planck Collaboration XI}(2016)}]{planck/11:2015}
{Planck Collaboration XI}. 2016, \aap, 594, A11

\bibitem[{{Planck Collaboration XIII}(2016)}]{planck/13:2015}
{Planck Collaboration XIII}. 2016, \aap, 594, A13

\bibitem[{{Planck Collaboration XIV}(2016)}]{planckXIV}
{Planck Collaboration XIV}. 2016, \aap, 594, A14

\bibitem[{{Planck Collaboration XVI}(2014)}]{planck/16:2013}
{Planck Collaboration XVI}. 2014, \aap, 571, A16


\bibitem[{{Riess} {et~al.}(2018){Riess}, {Casertano}, {Yuan}, {Macri},
  {Anderson}, {Mackenty}, {Bowers}, {Clubb}, {Filippenko}, {Jones}, \&
  {Tucker}}]{riess2018}
{Riess}, A.~G., {Casertano}, S., {Yuan}, W., {et~al.} 2018, \apj, 855, 136 

\bibitem[{Sellentin \& Heavens(2016)}]{Sellentin:2015waz}
Sellentin, E., \& Heavens, A.~F. 2016, Mon. Not. Roy. Astron. Soc., 456, L132

\bibitem[{{Sievers} {et~al.}(2013){Sievers}, {Hlozek}, {Nolta}, {Acquaviva},
  {Addison}, {Ade}, {Aguirre}, {Amiri}, {Appel}, {Barrientos}, {Battistelli},
  {Battaglia}, {Bond}, {Brown}, {Burger}, {Calabrese}, {Chervenak}, {Crichton},
  {Das}, {Devlin}, {Dicker}, {Bertrand Doriese}, {Dunkley}, {D{\"u}nner},
  {Essinger-Hileman}, {Faber}, {Fisher}, {Fowler}, {Gallardo}, {Gordon},
  {Gralla}, {Hajian}, {Halpern}, {Hasselfield}, {Hern{\'a}ndez-Monteagudo},
  {Hill}, {Hilton}, {Hilton}, {Hincks}, {Holtz}, {Huffenberger}, {Hughes},
  {Hughes}, {Infante}, {Irwin}, {Jacobson}, {Johnstone}, {Baptiste Juin},
  {Kaul}, {Klein}, {Kosowsky}, {Lau}, {Limon}, {Lin}, {Louis}, {Lupton},
  {Marriage}, {Marsden}, {Martocci}, {Mauskopf}, {McLaren}, {Menanteau},
  {Moodley}, {Moseley}, {Netterfield}, {Niemack}, {Page}, {Page}, {Parker},
  {Partridge}, {Plimpton}, {Quintana}, {Reese}, {Reid}, {Rojas}, {Sehgal},
  {Sherwin}, {Schmitt}, {Spergel}, {Staggs}, {Stryzak}, {Swetz}, {Switzer},
  {Thornton}, {Trac}, {Tucker}, {Uehara}, {Visnjic}, {Warne}, {Wilson},
  {Wollack}, {Zhao}, \& {Zunckel}}]{sievers/etal:2013}
{Sievers}, J.~L., {Hlozek}, R.~A., {Nolta}, M.~R., {et~al.} 2013, \jcap, 10, 60

\bibitem[{{Story} {et~al.}(2013){Story}, {Reichardt}, {Hou}, {Keisler}, {Aird},
  {Benson}, {Bleem}, {Carlstrom}, {Chang}, {Cho}, {Crawford}, {Crites}, {de
  Haan}, {Dobbs}, {Dudley}, {Follin}, {George}, {Halverson}, {Holder},
  {Holzapfel}, {Hoover}, {Hrubes}, {Joy}, {Knox}, {Lee}, {Leitch}, {Lueker},
  {Luong-Van}, {McMahon}, {Mehl}, {Meyer}, {Millea}, {Mohr}, {Montroy},
  {Padin}, {Plagge}, {Pryke}, {Ruhl}, {Sayre}, {Schaffer}, {Shaw}, {Shirokoff},
  {Spieler}, {Staniszewski}, {Stark}, {van Engelen}, {Vanderlinde}, {Vieira},
  {Williamson}, \& {Zahn}}]{story/etal:2013}
{Story}, K.~T., {Reichardt}, C.~L., {Hou}, Z., {et~al.} 2013, \apj, 779, 86

\bibitem[{Szapudi {et~al.}(2001)Szapudi, Prunet, Pogosyan, Szalay, \&
  Bond}]{PolSpice1}
Szapudi, I., Prunet, S., Pogosyan, D., Szalay, A.~S., \& Bond, J.~R. 2001, The
  Astrophysical Journal Letters, 548, L115

\bibitem[{{Weiland} {et~al.}(2018){Weiland}, {Osumi}, {Addison}, {Bennett},
  {Watts}, {Halpern}, \& {Hinshaw}}]{weiland2018}
{Weiland}, J.~L., {Osumi}, K., {Addison}, G.~E., {et~al.} 2018, \apj, 863, 161


\end{thebibliography}

%
%
%
%
%
%
%
%
%

\end{document}